\magnification\magstep1
\baselineskip=16pt

{\bf On the necessary conditions for the occurrence of the
"environment-in\-du\-ced superselection rules"}

\medskip

\centerline{Miroljub Dugi\' c}

\medskip

\centerline{\it Department of Physics, Faculty of Science}

\smallskip

\centerline{\it P.O.Box 60, 34 000
Kragujevac, FR Yugoslavia}

\bigskip

E-mail address : dugic@uis0.uis.kg.ac.yu

{\bf Abstract :} We briefly summarize the main recently obtained results
concerning existence of the (effective) necessary conditions for the
occurrence of the "environment-induced superselection rules" 
(decoherence).


There is considerable interest in the subject of decoherence in the
different arreas of modern physics.

Recently [1, 2], in the context of the "environment-induced superselection
rules (EI\-SR)" (or decoherence) theory [3, 4], it was pointed out 
existence of the (effective) necessary conditions for the occurrence
of the EI\-SR (decoherence) effect. Here we briefly summarize the main results
in this concern, and distinguish some lines of the research, which is still 
in progress.

We shall first precisely present the methodological basis and the statements
concerning the necessary conditions. Later, we briefly discuss the physical 
meaning of the results, with an emphasis on the question of universality
of the EI\-SR effect.

The EISR theory [3,4] relies upon the, so-called, "orthodox method",
i.e. upon the formalism of the unitary operator of evolution in time,
$\hat U(t)$, of the composite quantum system, "(open) system plus environment
(S+E)". More precisely, if $\hat \rho_{S+E}(t=0)$ represents the initial
state of the composite system, then $\hat \rho_{S+E}(t)$ reads :
$$\hat \rho_{S+E}(t) = \hat U(t) \hat \rho_{S+E}(t=0) \hat U^{\dag}(t),
\eqno (1)$$

\noindent
while the "tracing out" defines the (sub)system's "density matrix" 
$\hat \rho_S(t)$ :
$$\hat \rho_S(t)  = tr_E \hat \rho_{S+E}(t), \eqno (2)$$

The method is the very basis of all the proper methods in the field (such
as the "path integration technique", and the "master equations" formalism),
and bears model-independence. This fact also follows from the analysis
by Zurek [3] which distinguishes that, effectively, all the calculations
refer to the spectral form of the interaction Hamiltonian of the composite
system, $\hat H_{int}$.

On the basis of the next set of assumptions : (i)  $\hat H_{int}$ 
is time independent, (ii)  $\hat H_{int} = \sum_{m,n} \gamma_{mn}
\hat P_{Sm} \otimes \hat \Pi_{En}$ (where appear the projectors onto the
subspaces of the Hilbert space of the system, $H^{(S)}$,
and of environment, $H^{(E)}$) - which 
is [1, 2] refered to as the separability of  $\hat H_{int}$, and (iii)
Initial state of the environment is normalizable one, Zurek [3] has
obtained the next results, which have been used [1, 2] as a {\it formal
definition} of the EISR effect. The results of Zurek
are as folows. In a basis $\{\vert m\rangle \}$ of $H^{(S)}$, which is 
adapted to the decomposition of $H^{(S)}$ defined by the 
projectors $\hat P_{Sm}$ in the above point (ii) - which
is refered to as the "pointer basis" [3] - one obtains:
$$\rho_{Smm'} = C_m C^{\ast}_{m'} z_{mm'}(t), \eqno (3)$$

\noindent
where the correlation amplitude, $z_{mm'}(t)$, for each $m \neq m'$
satisfies :
$${\rm (a)} \lim_{t \to \infty} \lim_{N \to \infty} z_{mm'}(t) = 0,$$
$${\rm (b)} \lim_{t \to \infty} \langle z_{mm'}(t)\rangle_t = 0,$$
$${\rm (c)} \lim_{t \to \infty} \Delta z_{mm'}(t) \propto N^{-1/2}$$

\noindent
where N-being a number proportional to the number of particles in 
the environment, and $\Delta z_{mm'}(t)$-being the "standard deviation" 
of $z_{mm'}(t)$.

In his recent paper [4], Zurek has introduced another element of 
definition of the EISR effect :
$${\rm (d)} \hat U(t) \vert m\rangle_S \otimes \vert \chi\rangle_E =
\vert m\rangle_S \otimes \vert chi_{m}(t)\rangle_E,$$

\noindent
for arbitrary initial state of the environement, $\vert \chi\rangle_E$.

The condition (d) is a formal expression of the requirement [4]
for stability of the "preferred set of states". Since the "preferred set"
needs not to consist in mutually exactly orthogonal states, the above
point (d) represents a slight idealization, but which can be derived 
(cf. Appendix II of Ref. [2]) from the original statement.

Bearing in mind the formal definition of decoherence - i.e., the above
points (a)-(d) - the occurrence of the EISR effect has been investigated
[1, 2] by adopting the assumptions oposite to the above assumptions
(i)-(iii). Since the assumptions underlying the defining points (a)-(d) have
all been adopted as a part of the definition, 
the results to be presented below refer exactly, but not necessarily
exclusively, to the EISR theory [3, 4].

The main result of the analysis [1, 2] is existence of the (effective)
{\it necessary conditions} for the occurrence of the EISR effect.
These conditions are as follows: (A) Diagonalizability of $\hat H_{int}$
in a noncorrelated basis of the Hilbert state space of the composite 
system - which is refered to as separability of $\hat H_{int}$, and
(B) "Nondemolition" character of $\hat H_{int}$ : $[\hat H_{int}(t), 
\hat H_{int}(t')] = 0$.

Effectiveness of the conditions refers to the cases in which the 
necessary conditions are not fulfilled, but for which the points (a)-(d)
prove justified. In the context of the EISR theory these cases represent the 
pathological cases. In a (hypothetical) wider theory (which, by
definition, should involve the original EISR theory) these cases appear as
the particular exceptions of the next rule :

(R1) Whenever the necessary conditions are not satisfied, the EISR
effect does not occur.
[Note : existence of the "exceptions" has not been proved, but is just
not forbidden.]

The necessary conditions have a simple physical meaning : they represent
the conditions of {\it existence} of the "pointer basis" of the open system 
$S$. In other words: apart from the possible exceptions, if any of the
above conditions ((A) or (B)) is not satisfied, the pointer basis of the
open system does not exist. And this should not be missed with the
cases in which pointer basis exists, but for which EISR does not occur
due to some details [1] in the model ; this essentially refers to 
nonexistence of the sufficient conditions of EISR.

The necessary conditions directly refer to the question of universality
of the EISR effect. On the other side, the question of universality seems
to be fundamental in the context of the problem of "transition from
quantum to classical" [5].

It seems that in the context of the EISR theory [3, 4], 
the universality is an 
implicit and plausible statement, which can be expressed by the next rule :

(R2) Appart from the "technical" details, for the realistic physical
models, whenever the system is in unavoidable
interaction with its environment, there occurs the EISR effect.

The rule (R2) is a statement of the original theory [3-5]. The rule (R1)
is a result of the recent progress [1, 2] in the field. It is apparent that,
now, the rule (R1) makes the rule (R2) essentially irrelevant, and
puts a limit on the universality of the EISR effect.

This brings us to the questions and problems which follow on the basis 
of the rule (R1). Here we just distingusih the main operational tasks,
likewise the problem of deeper physical meaning of the rule (R1).

As the main operational task in this concern appears the task of
investigating separability (and/or "nondemolition" character) of
$\hat H_{int}$. The investigation refers to the existing models,
and, in addition, points to the criteria for modeling the interaction
Hamiltonians in the context of the quantum measurement, and the EISR
theory; for an example of the later see Ref. [6].

On the other side, the problem of deeper physical meaning and 
consequences of the rule (R1) calls for making the different
strategies and the corresponding research programms, 

and represents an investigation which is still in progress.

We conclude that existence of the (effective) necessary conditions
for the occurrence of the EISR effect opens some questions in the field.
In a hypothetical wider-EISR theory, the conditions lead to establishing 
the new rule, (R1), of the theory, and puts speciffic limit concerning
universality of the EISR effect. Further
researche is expected to involve both, dealing with some operational
tasks (such as investigating separability of $\hat H_{int}$), likewise
investigating the deeper physical meaning and consequences of the
necessary conditions.

{\bf References :}

\item{[1]}
M. Dugi\' c, Physica Scripta {\bf 53} (1996) 9

\item{[2]}
M. Dugi\' c, Physica Scripta {\bf 56} (1997) 560

\item{[3]}
W. H. Zurek, Phys. Rev. {\bf D26} (1982) 1862

\item{[4]}
W. H. Zurek, Prog. Theor. Phys. {\bf 89} (1993) 281

\item{[5]}
W. H. Zurek, Phys. Today, October 1991, p. 26

\item{[6]}
M. Dugi\' c, Phys. Lett. {\bf A235} (1997) 200

\end